\documentclass[12pt,a4paper,dvips]{article}
\usepackage{graphicx}
\usepackage{amstext}
\usepackage{amsmath}
\usepackage{subfigure}
\usepackage{l3_title,ifthen}
\usepackage{cite,mcite}
\usepackage{endfig}
\journalname{Phys. Lett. B}
\date{July 16, 2004}
\preprint{2004-023}      
\def\TeV{\ifmmode {\mathrm{\ Te\kern -0.1em V}}\else
                   \textrm{Te\kern -0.1em V}\fi}%
\def\MeV{\ifmmode {\mathrm{\ Me\kern -0.1em V}}\else
                   \textrm{Me\kern -0.1em V}\fi}%
\def\GeV{\ifmmode {\mathrm{\ Ge\kern -0.1em V}}\else
                   \textrm{Ge\kern -0.1em V}\fi}%
\def\Zo{\ensuremath{\mathrm {Z}}}
\def\epem{\ensuremath{\mathrm{e^+e^-}}}%
\def\mm{\ensuremath{\mathrm{\mu^+ \mu^-}}}%

\newcommand{\tlo}[1]{\raisebox{-1.5ex}[1.5ex]{#1}}
\newcommand{\thi}[1]{\raisebox{1.5ex}[-1.5ex]{#1}}
\newcommand{\fluxtabHone}{&&&&&&&&&}
\newcommand{\fluxtabFone}{\multicolumn{2}{|c|}{\tiny\thi{momentum}} & 
                           \tiny $\langle p \rangle$                & 
                           \tiny$\Phi\cdot\langle p \rangle^3$      & 
                           \tiny$\Delta_\Phi^\mathrm{stat}$         & 
                           \tiny\tlo{$\rho_\Phi$}                   & 
                           \tiny$\Delta_\Phi^\mathrm{syst}$}          
\newcommand{\fluxtabRone}{ \tiny$\tlo{R}$                           & 
                           \tiny$\Delta_R^\mathrm{stat}$            & 
                           \tiny\tlo{$\rho_R$}                      & 
                           \tiny$\Delta_R^\mathrm{syst}$}             
\newcommand{\fluxtabFtwo}{\multicolumn{2}{|c|}{\tiny \thi{interval}}& 
                           \tiny[\GeV{}]                            & 
                           \tiny$\left[\frac{\mathrm{\GeV{}}^2}	      
                                 {\mathrm{cm}^2\mathrm{s}\,	      
                                 \mathrm{sr}}\right]$               & 
                           \tiny[\%]                                & 
                                                                    & 
                           \tiny[\%]}				      

\newcommand{\fluxtabRtwo}{ \tiny                                    & 
                           \tiny  [\%]                              & 
                           \tiny                                    & 
                           \tiny   [\%]}			      
\newcommand{\fluxtabHtwo}{\multicolumn{2}{|c|}{\tiny  \thi{$[$\GeV{}$]$}}}

\begin{document}
\begin{titlepage}
\title{Measurement of the Atmospheric Muon Spectrum from 20 to 3000~\GeV{}}
\author{The L3 Collaboration}

%
\begin{abstract}
The absolute muon flux between 20 \GeV{} and 3000 \GeV{} is measured with
the  L3 magnetic muon spectrometer for zenith angles  ranging from 
$\rm 0^{\circ}$ to $\rm 58^{\circ}$. Due to the large exposure of about 150~m$^2$\,sr\,d, 
and the excellent momentum resolution of the L3 muon chambers,
a precision of 2.3\,\% at 150~\GeV{} in the vertical direction is achieved.

The ratio of positive to negative muons is studied between 20~\GeV{} and 
500~\GeV{}, and the average vertical muon charge ratio 
is found to be 1.285 $\pm$ 0.003 (stat.) $\pm$ 0.019 (syst.).
\end{abstract}
\vspace{0.5cm}

{\it The L3+C group dedicates this publication to the late Bianca Monteleoni.}

\submitted
\end{titlepage}

%
%
\section*{Introduction}
Atmospheric muons are among the final products of cosmic ray induced
air-showers. The absolute muon flux and its
momentum dependence
are mainly determined by the  flux of nucleons entering the 
atmosphere and the inclusive meson production cross sections in
high-energy hadronic interactions. The ratio of the fluxes of
positive to negative muons, denoted as charge ratio in the following,
reflects the proton to neutron ratio at the top of the atmosphere,
folded with the production and decay spectra of charged pions and kaons.
While the knowledge of the primary cosmic ray spectrum below a few 100~\GeV{} has 
improved considerably in the recent past 
\cite{cospr:compil}, large uncertainties still exist in the primary energy 
range between 0.1~\TeV{} and 500~\TeV{} responsible for the production of secondaries with momenta in the range 
under study here. Moreover, the details of high 
energy hadronic interactions still lack theoretical understanding
and there is little experimental data in the relevant energy and phase space regions~\cite{cosmo:engel}.
Therefore the ground-level muon flux and charge ratio are widely used to
tune or verify the parameters of atmospheric cascade 
calculations~\cite{cosnu:honda,cosnu:agrw,mucalc:bugaev,cosnu:vadim}. 
Currently these calculations are of great interest, as they 
predict the absolute atmospheric neutrino fluxes~\cite{cosnu:gaihon} 
which are needed to interpret the observed muon neutrino flux 
deficit~\cite{cosnu:SK98,cosnu:Sou97,cosnu:MACRO98} 
and to evaluate the backgrounds for neutrino astronomy.

The muon flux and charge ratio have been extensively studied
with different experimental techniques~\cite{cosmu:thct}. 
However, results show discrepancies of about 10$-$20\,\% with respect 
to each other, which exceed the uncertainties assigned to the individual 
measurements and thus indicate the presence of systematic effects
not accounted for. 

Here a new measurement of the atmospheric muon flux is presented
using the precise muon spectrometer of the L3 detector located
at the LEP collider at CERN, near Geneva, Switzerland. 
Special attention is given to the precise determination of all relevant
detector and environmental parameters needed to convert the raw-data distributions into 
an absolute surface level flux. The large statistics available permits
extensive studies of the residual systematic uncertainties.
%
%
\section*{Experimental setup}
\begin{figure}[!t]
     \includegraphics[clip,width=\linewidth]{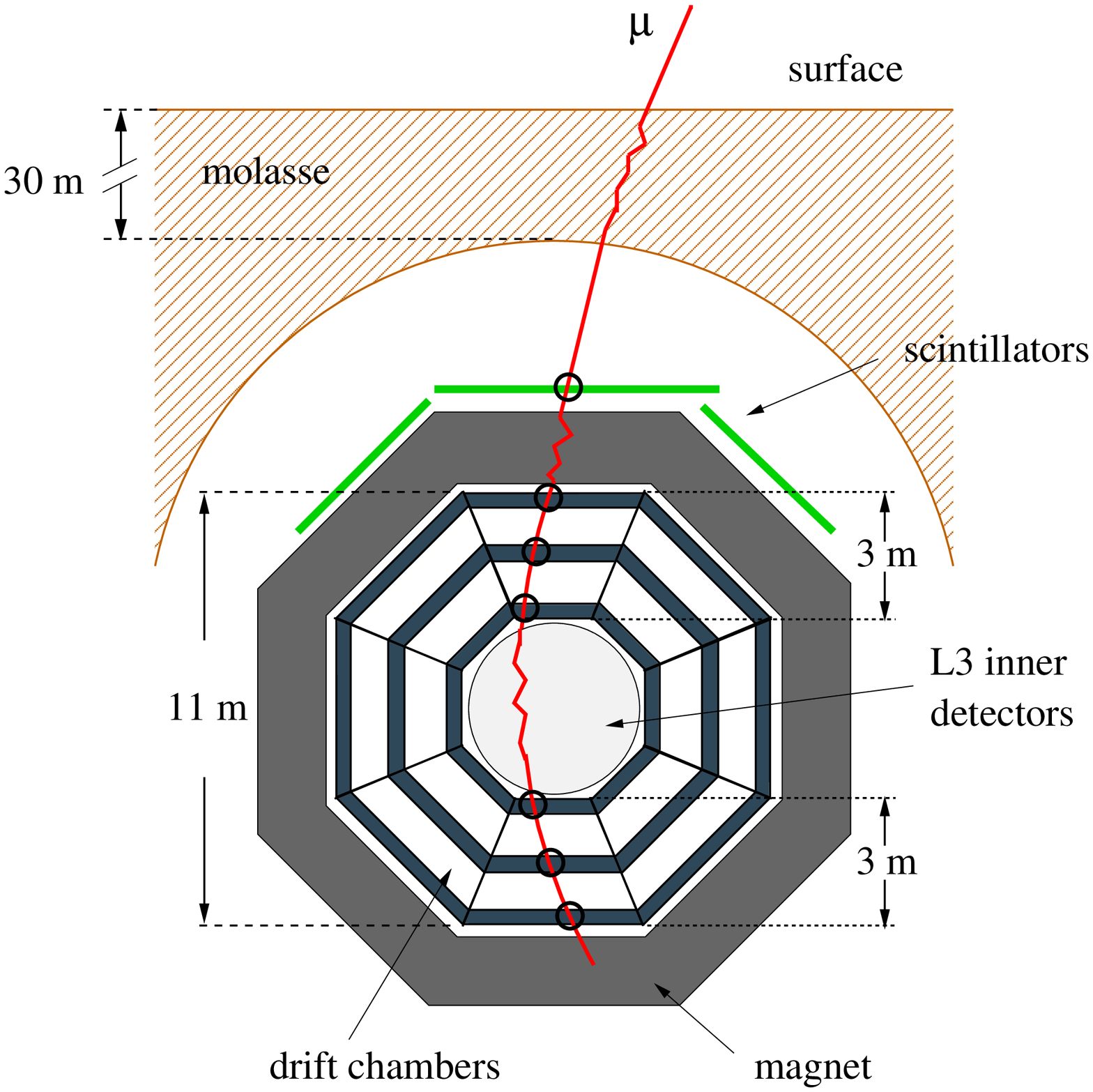}
     \caption{{Schematic view of the experimental setup}}
   \label{fig:l3c}
\end{figure}

The momentum distribution of atmospheric muons is measured 
with the upgraded L3 setup of the L3 detector~\cite{det:l3} known as L3+C~\cite{det:l3cnim}. The parts
of the detector used in this analysis are sketched
in Figure~\ref{fig:l3c}. 
After passing through the stratified rock overburden, called "molasse", 
the arrival time $t_0$ of a muon is measured with a resolution of 1.7~ns by
a 202~m$^2$ scintillator array placed on top of the L3 detector. The array is
composed of 34 modules, each read out by two photomultipliers in coincidence
to reduce noise.
Inside a volume of about 1000~m$^3$ with a magnetic field 
of 0.5~T,  the coordinates and slopes of a muon track are measured in up to six 
drift chambers in the bending plane and up to eight times in the non-bending plane.
These chambers
are arranged concentrically around the LEP beam line in two groups of eight octants, 
each containing three layers of drift cells. By subtracting the $t_0$ time 
from the arrival times of the drift electrons at the sense wires, a track position
in each chamber 
can be reconstructed with a precision of about 60~$\mu$m in the bending plane and 1~mm 
in the non-bending plane.

Only three points are needed to determine the radius
of the track in the magnetic field, therefore the momentum of 
a muon traversing
two octants can be measured twice. This redundancy is used to
evaluate the detector efficiencies and the resolution of the apparatus.
The best resolution is obtained when fitting the six points together over
the full track length of 11~meters. The multiple scattering and energy loss
inside the L3 inner detectors, as well as the effect of the inhomogeneous magnetic field
are taken into account using the procedure proposed in Reference~\cite{soft:GEANEfit}.

Equipped with a trigger and data acquisition system independent of 
the normal L3 data-taking, L3+C recorded $\rm 1.2 \times 10^{10}$ atmospheric muon
triggers during its operation in the years 1999 and 2000. 

The L3+C experiment was located 450~m above sea level at a longitude of
 6.02$^{\circ}$~E and a latitude of 46.25$^{\circ}$~N.
 
For vertically incident muons, the mean energy loss in the molasse 
($X$= 6854~g\,cm$^{-2}$) and the magnet ($X$=1227~g\,cm$^{-2}$) is 19~\GeV{}  
at low momenta and reaches 57~\GeV{} at 1~\TeV{}. 
%
%
\section*{Analysis}
\label{sec:analysis}
     %
     %
\subsection*{Detector and molasse simulation}
The geometrical acceptance of the L3+C detector and the
stochastic energy loss in the molasse overburden are evaluated using the following 
simulation procedure: Monte Carlo events are generated on the surface using a 
parameterization of the zenith angle and momentum dependence of the muon spectrum 
as obtained with the \texttt{CORSIKA}~\cite{soft:CORSIKA} program. 
These simulated muons are then tracked through a 
\texttt{GEANT}~\cite{soft:GEANT,soft:GEANTpatch} model of the L3+C environment 
which includes the molasse, access shafts and the concrete structures around 
the cavern which hosts the apparatus. Finally, the detector response is 
simulated with a detailed \texttt{GEANT} description of the L3 detector. 
The generated detector signals are reconstructed with 
the same program used for the data. In total $\rm 1.7 \times 10^9$ reconstructed 
Monte Carlo events are used
in this analysis.
     %
     %
\subsection*{Event selection}
\label{sec:evsel}
The data analysis is restricted to events with
three position measurements in at least one octant, a scintillator hit
and good running conditions during data-taking.
A total of $\rm 1.2 \times 10^9$ reconstructed muon tracks are retained.

The shielding of the 30~m of molasse overburden absorbs most of 
the charged air-shower particles other than muons. 
The number of muons produced in $ \rm e^+e^-$ collisions by LEP is negligible
compared to the flux of atmospheric muons. Therefore no background rejection
is needed. The data selection focuses on two topics. Firstly, fiducial 
volume cuts are defined to assure a good description of the data by the simulation.
Secondly, selection cuts are imposed on the track quality to enhance the momentum 
and angular resolution. These selection criteria are:
\begin{itemize}
\item The muon track positions must be measured in six layers
 in the bending plane.
\item The momentum resolution, calculated from the quality of the track position
measurements, should not exceed its nominal value by
more than 50\,\%. 
\item At least four position measurements (two in each octant)
 should be present in the non-bending plane.
\item The $\rm \chi^{2}$ of a fit of the tracks to a circle within an octant must satisfy $\rm \chi^{2} / ndf \, < \, 4$.
\item The difference between the two photomultiplier time measurements from the same 
scintillator module must be below 8 ns.
\end{itemize}
After these cuts, $ \rm 2 \times 10^7$ data events remain for the 
muon spectrum analysis. \\
The selection efficiencies depend on the charge, momentum and direction of the muon. For
muons with momenta above 80~\GeV, the average efficiency is 7.6\,\% for the fiducial volume cut and
33.3\,\%  for the quality selection.\\
4 \% of the raw events are multi-muon events. Each individual
    muon is counted as an input to the spectrum data.
     %
     %
\subsection*{Momentum resolution}
\begin{figure}[!t]
   \begin{center}
     \includegraphics[width=.96\linewidth]{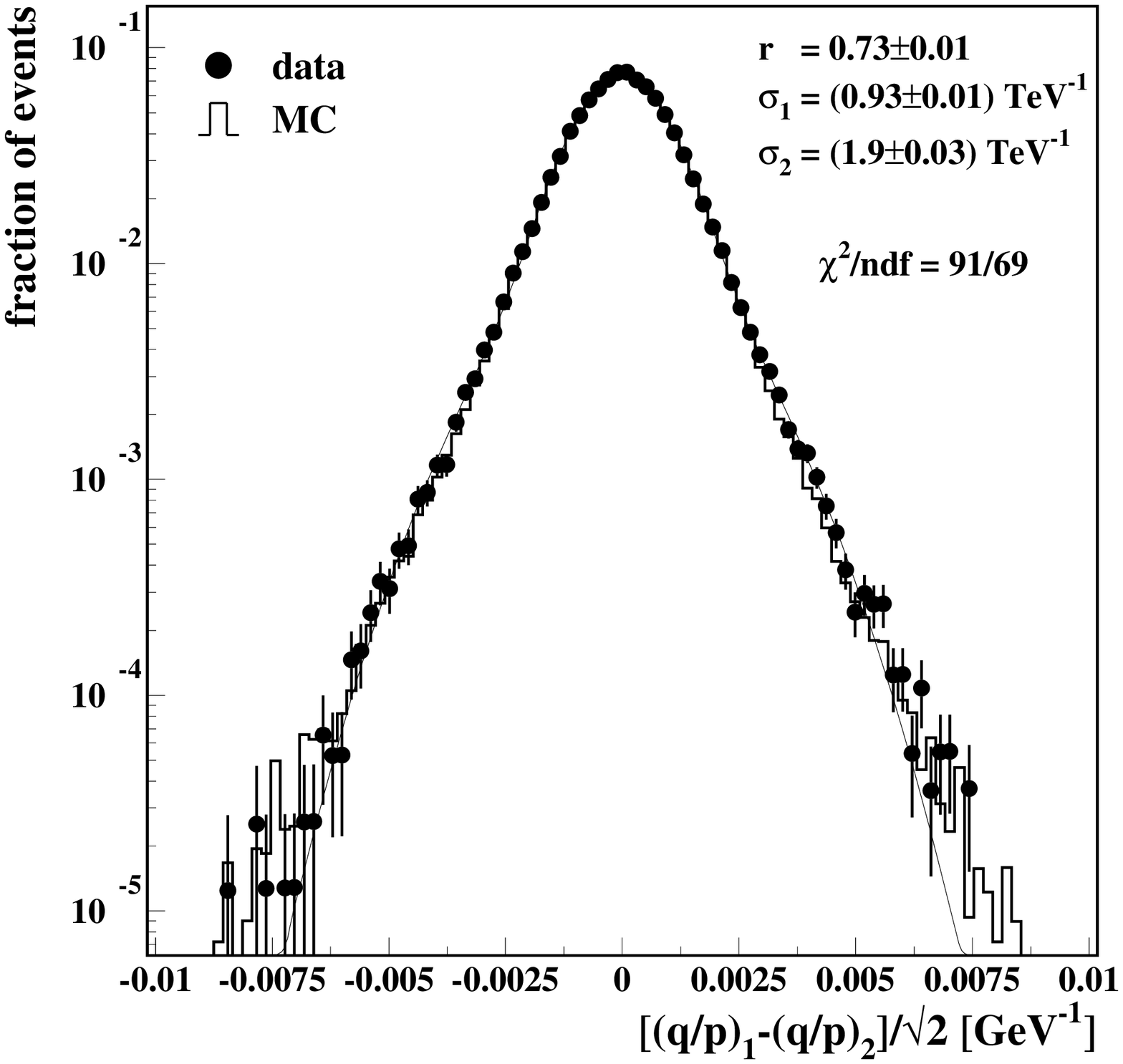}
     \caption{{Curvature difference at 100~\GeV. The line denotes a fit 
with a sum of two Gaussian distributions with width $\sigma_1$ and $\sigma_2$. The fraction of 
events with width $\sigma_1$ is denoted by $r$.}}
   \label{fig:resoexa}
   \end{center}
\end{figure}

\begin{figure}[!t]
   \begin{center}
     \includegraphics[width=.96\linewidth]{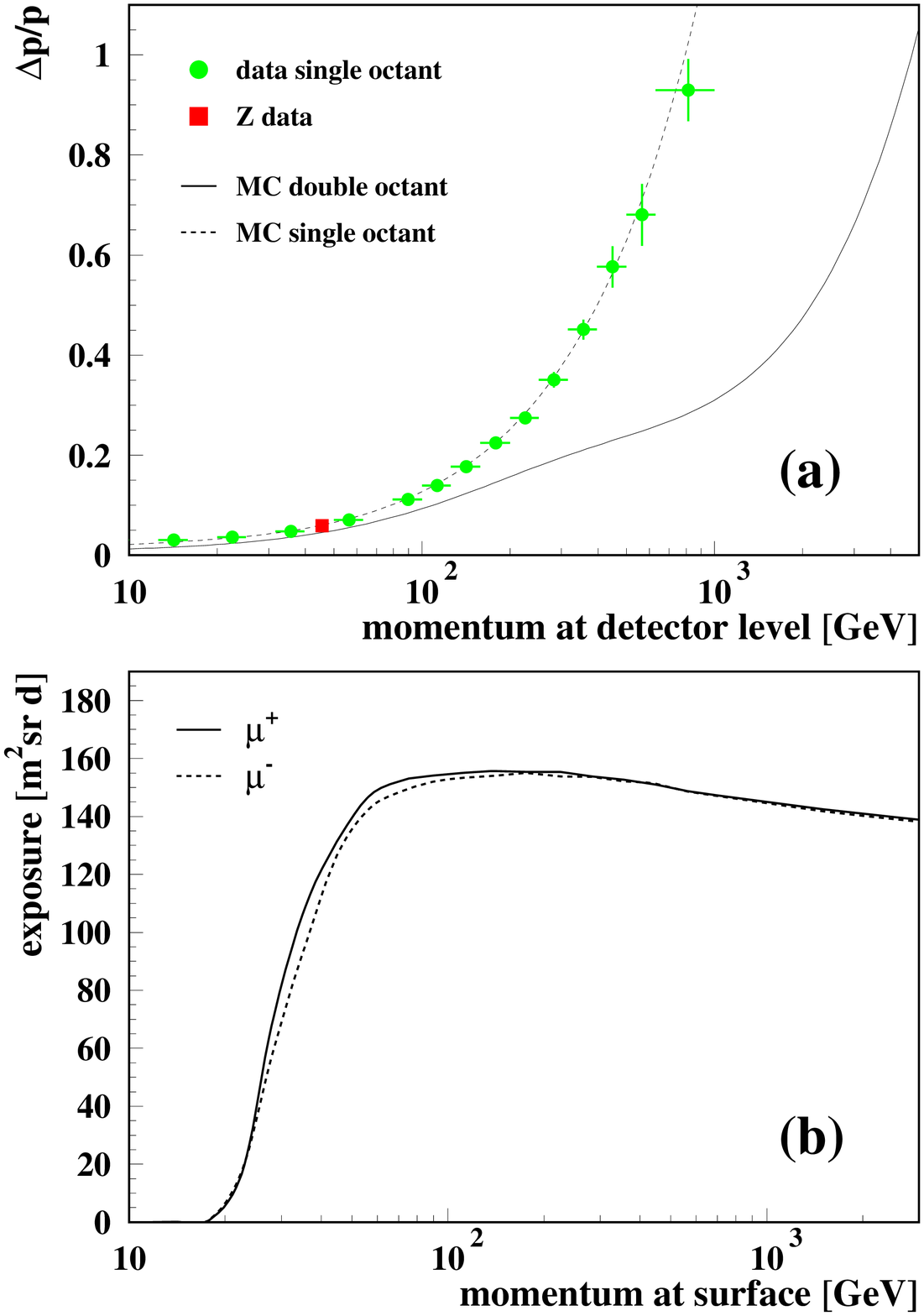}
     \caption{{L3+C detector performance: (a) relative momentum resolution as a function of the muon momentum at the detector-level, (b) detector exposure for this analysis as a function of the muon momentum at surface for 
                positive and negative muons (the sum over all zenith angle bins is shown).}}
   \label{fig:reso}
   \end{center}
\end{figure}

The single-octant resolution is inferred directly from the data by comparing
the two independent curvature measurements of muons
traversing two octants. An example of the curvature difference is shown in Figure~\ref{fig:resoexa}.
These measurements are used to tune the detector simulation,
from which the resolution of the full fit is determined.
The relative momentum resolution $\Delta p/p$ as a function of 
momentum at the detector-level
is shown in Figure~\ref{fig:reso}(a). 
The maximum detectable momentum of the spectrometer, defined as the 
momentum at which $\Delta p/p$ reaches unity, is  0.78~\TeV{} for muons measured
in only one octant and about 5~\TeV{} for muons measured in two octants.

     %
     %
\subsection*{Detector efficiencies}
The efficiency of each subdetector is studied by exploiting redundancies 
in the measurement process. For about 50\,\% of the tracks, the muon arrival
time is also deduced from the muon chambers.
 These tracks are used to determine the scintillator efficiencies
as a function of time and position on the array. A mean efficiency of  95.6\,\% is found 
at the start of data-taking decreasing continuously to 94.5\,\% towards the end of 2000. 
The possibility of reconstructing a muon within a single octant is used to scan the drift-layer 
performance of the facing octant. On average, a fraction of 10.5\,\% of the drift cells 
are found  to have an efficiency lower than 80\,\%. These regions are excluded in both the data 
and Monte Carlo reconstruction. Under these conditions the trigger efficiency
is determined from redundant trigger classes to be 99.85\,\% on average.

During data-taking, the total effective running time was continuously
 measured with a 10~MHz live-time counter, which
is disabled whenever the trigger system is not ready to accept new data. In addition, each 
second an external trigger signal was sent to the L3+C trigger system. The
number of these external triggers on tape compared to the total number of running
seconds gives another estimate of the effective running time and agrees with the 
value from the live-time counter within 0.02\,\%.
     %
     %
\subsection*{Selection efficiency}
Using the possibility to
measure a muon independently in two detector parts, the selection efficiencies
are determined in the following way: the detector is subdivided into two hemispheres, $i$ and $j$,
and the conditional hemisphere selection probabilities
$\varepsilon_i$
are measured for data and Monte Carlo separately as a function of the muon charge $q$, momentum 
$p$ and zenith angle $\theta$. In the absence of correlated inefficiencies,
the total selection efficiency for accepting a track in the two hemispheres is given
by the product $\varepsilon_1\times\varepsilon_2$. The ratio
\begin{equation}
    r=\frac{(\varepsilon_1\times\varepsilon_2)^\mathrm{data}}
    {(\varepsilon_1\times\varepsilon_2)^\mathrm{MC}}
\end{equation}
is used to correct the differences between data and Monte Carlo. Depending
on the zenith angle range and the data-taking year, 
$r$ varies from 0.84 to 0.90. A large fraction of this correction
factor originates from a defect in the TDCs used to read out the muon chambers, 
giving rise to an 8\,\% inefficiency for the full track selection.
     %
     %
\subsection*{Surface spectrum}
The relation between the momentum distribution measured in L3+C and
the muon surface spectrum is given by
\begin{equation}
    \mathbf n = \tau \cdot \mathbf {E\cdot R\cdot A\cdot m} \,.
    \label{eq:spec}
\end{equation}
Here $\mathbf n$ is the vector of events $n_i$ with 
measured momenta between $[{qp}_i,{qp}_{i+1}]$. The effective 
live-time is given by $\tau$ and
$\mathbf R$ denotes the migration matrix, {\it i.e.} the conditional 
probability
of measuring a momentum  ${qp}_i$ given a surface momentum $q{p}_j$. $\mathbf A$ is
the diagonal matrix of geometrical acceptances as a function of the  
surface momentum and $\mathbf E$ is the diagonal matrix of detector efficiencies
as a function of momentum at the detector-level.
The vector $\mathbf m$ contains
the true surface spectrum integrated over a surface momentum bin.
The complete detector matrix, $\mathbf D \equiv \mathbf {E \cdot R\cdot A} $,
is evaluated from the measured detector efficiencies and the detector simulation 
as follows:
\begin{equation}
    D_{ij} = \varepsilon_i \,r_i\, S_{\text{MC}}\, 
    \left( \frac{n_{ij}^{\text{sel}}}{N_j^{\text{gen}}}\right)_{\text{MC}} \Delta\Omega\;,
    \label{eq:detector}
\end{equation}
where $S_{\text{MC}}$ is the surface area used in the Monte Carlo generator,
$\Delta\Omega$ the solid angle
of the zenith bin under study, $\varepsilon_i$ includes the scintillator and
trigger efficiencies and $r_i$ is the selection efficiency correction
discussed above. $n_{ij}^{\text{sel}}$ denotes the number of selected Monte Carlo events 
found within a  detector-level momentum bin $i$, which were generated within the momentum 
bin $j$ at the surface, and $N_j^{\text{gen}}$ is the total number of Monte Carlo events 
generated within this surface momentum bin.

The effective
acceptance of this analysis is calculated by summing over the columns of the 
detector matrix, which yields the geometrical factor for a muon 
being registered in the detector and fulfilling the selection cuts. 
The product of the effective acceptance and the live-time gives the total exposure, 
shown in Figure~\ref{fig:reso}(b)
for positive and negative muons as a function of surface momentum.  
It rapidly decreases at low energies due to the momentum cut-off caused by
the molasse overburden. Below 200~\GeV{}, positive and negative muons have different
acceptances, because the magnetic field bends their tracks in opposite directions and
correspondingly into different detector regions. At large momenta the acceptance 
decreases with the performance of the full detector fit and
a more difficult reconstruction caused by the increasing production of delta rays.

The measurement Equation~(\ref{eq:spec}) is solved using the  least squares method,
 by minimizing
\begin{equation}
    \chi^2=\sum_i \frac{(n_i-\tau \sum_j D_{ij} m_j )^2}{\sigma_i(\mathbf{m})^2}\;,
    \label{eq:lsqchi2}
\end{equation}
where $\sigma_i$ contains the statistical errors of the data and the detector matrix:
\begin{equation}
     \sigma_i=\sqrt{n_i+\tau^2 \sum_j V[D_{ij}]\, m_j^2}\;.
     \label{eq:newerr}
\end{equation}
In the first step, the statistical Monte Carlo variances $V[D_{ij}]$ are set to zero,
such that Equation~(\ref{eq:lsqchi2}) becomes linear with respect to the surface spectrum~$\mathbf{m}$
and its solution is
\begin{equation}
    \mathbf{\widehat{m}} = \frac{1}{\tau} \;\left(
    \mathbf{D}^T \mathbf{W} \mathbf{D} \right)^{-1}
    \mathbf{D}^T \mathbf{W}  \mathbf{n}
    \label{eq:lsq}
\end{equation}
with covariance matrix
\begin{equation}
    \mathbf{V}[\mathbf{\widehat{m}}]=\left(\mathbf{D}^T \mathbf{W} \mathbf{D}\right)^{-1}\,.
    \label{eq:Vm}
\end{equation}
Here $\mathbf{W}$ denotes the diagonal weight
matrix containing the statistical errors of the data and the Monte Carlo, 
$W_{ii}=1/\sigma_i^2$.

The minimization of Equation~(\ref{eq:lsqchi2}) is then repeated using the solution 
$\mathbf{\widehat{m}}$ of the previous iteration for the calculation 
of the errors in~Equation~(\ref{eq:newerr}). This process is repeated until the maximum 
relative difference to the result of the previous iteration is below $10^{-6}$. 
Typically four iterations are needed.
%
%
\section*{Systematic uncertainties}
    %
    %
\subsection*{Normalization uncertainties}
Uncertainties on the live-time and the trigger and scintillator efficiencies 
give rise to a normalization uncertainty of 0.7\,\%. 

The uncertainty of
the detector acceptance is assessed in three studies: First, the results
obtained for statistically independent data subsamples, as for instance the data
collected in 1999 and 2000 or in different detector parts, are compared. 
Second, the muon flux and charge ratio are measured as a function of the azimuthal
angle. At large momenta, geomagnetic effects and the variation of the molasse overburden 
are not important, and therefore a flat distribution is expected. 
Finally, the stability of the measured 
flux and charge ratio with respect to a variation of the selection criteria is investigated.

From these studies, additional normalization uncertainties in the absolute muon flux are derived. These
range from 1.7\,\% to 3.7\,\% depending on the zenith angle.
For the charge ratio normalization uncertainties between 1.0\,\% and 2.3\,\% are estimated.

\begin{figure}[t!]
\begin{center}
       \includegraphics[width=\linewidth] {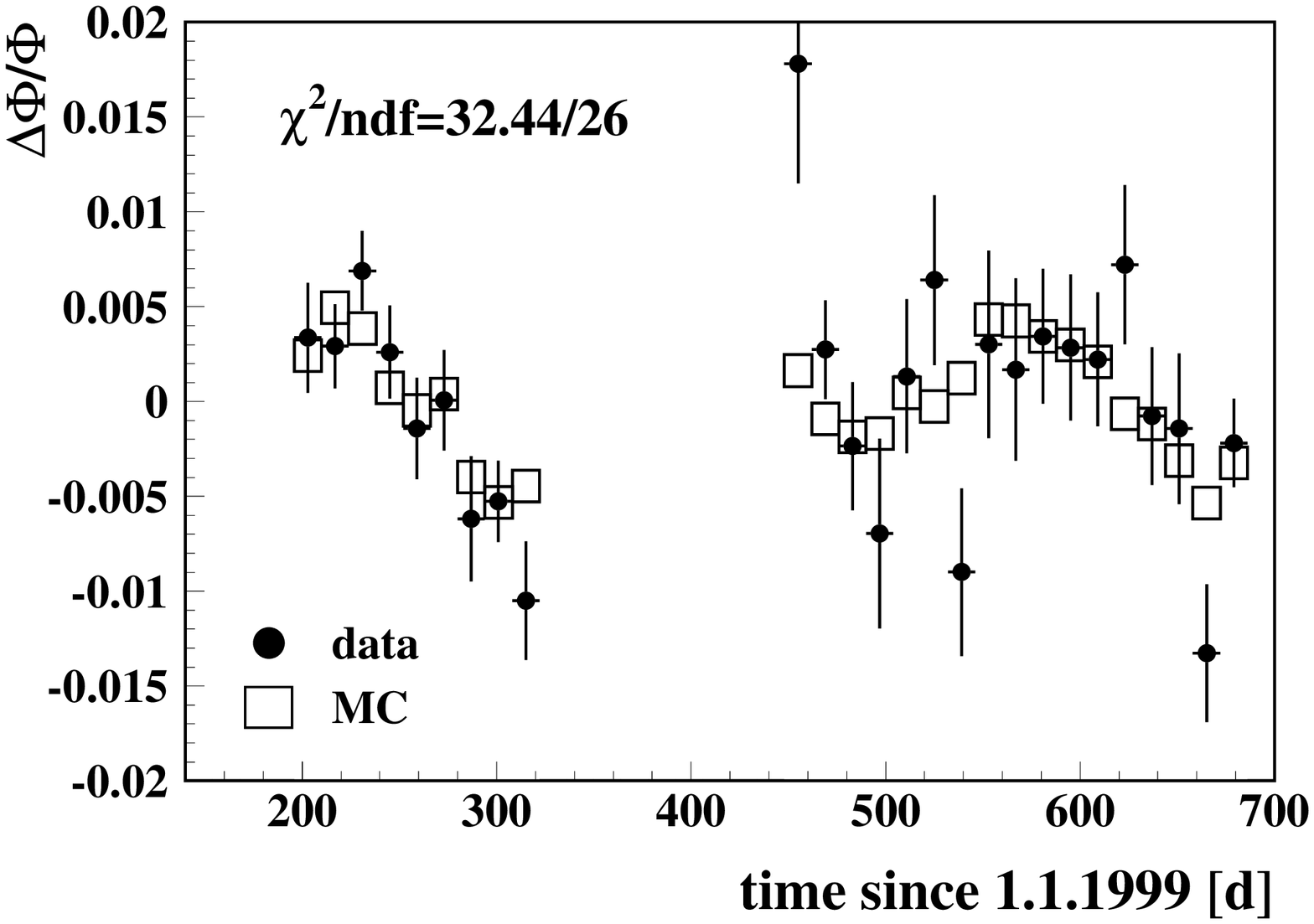}
        \caption{{Relative rate change with time for muon momenta between 50 and 62~\GeV{}, compared to 
                   a prediction of the atmospheric effect obtained with the \texttt{TARGET} air-shower 
                   simulation. 
                   The value of a $\chi^2$ comparison of data and Monte Carlo is also shown.}}
        \label{fit:rvst}
\end{center}
\end{figure}

Figure~\ref{fit:rvst} shows an example of the stability of the measured muon flux 
within the two data-taking years for zenith angles between $0$ and $32$~degrees.
Since the muon production is known to change with atmospheric conditions~\cite{atmo:barret,atmo:sagi}, 
the observed variation of the muon flux is compared to an air-shower simulation with the 
\texttt{TARGET}~\cite{soft:target} program using atmospheric density profiles measured in balloon 
flights close to the experiment~\cite{l3c:atprof}. Good overall agreement between data and Monte 
Carlo is observed. However, 
the full comparison to the rates in~26 weeks and 14~momentum bins
yields a $\chi^2/\mathrm{ndf}$ of 526/364. The assumption that this large value is caused by 
detector inefficiencies not accounted for, leads to an additional normalization uncertainty of 0.3\,\%, 
which is well within the above estimated uncertainties.\\
    %
    %
\subsection*{Momentum scale uncertainties}
Due to the steepness of the muon spectrum, even small uncertainties in the absolute
momentum scale can introduce a considerable bias in the muon flux measurement.

The uncertainty on the L3 magnetic field strength introduces a momentum 
scale bias of less than 0.4\,\%~\cite{det:l3b}.

Furthermore, the momentum measurement is subject to uncertainties of the detector alignment. 
A systematic shift of the chamber positions may introduce a 
constant offset $C$. The measurement of the curvature, $q/p$, and the alignment related
momentum scale uncertainty, $\delta_\mathrm{al}$ is given by
\begin{equation}
    \delta_\mathrm{al} = \frac{\Delta C} {q/p+\Delta C}\,p\; ,
\end{equation}
and depends on the muon charge. 
Within one octant, the alignment is measured by an optical alignment system~\cite{det:rasnik}
with a precision corresponding to 0.19~\TeV{}$^{-1}$~\cite{thes:fabre}. The relative
alignment of the muon chamber octants, relevant for this analysis, is determined from
the data itself with a precision between 0.075 and 0.152~\TeV{}$^{-1}$~\cite{thes:unger}, 
depending on the zenith angle.

The uncertainty due the molasse overburden  affects the conversion of 
the measured flux at the
detector to the surface. The results of two survey drillings at different locations close to L3+C
provide an absolute measurement of the L3+C matter overburden.
The influence of molasse inhomogeneities and of surface installations not
included in the L3+C simulation is estimated by studying the variance of the
muon flux as a function of the azimuthal angle near the momentum threshold. This leads
to an uncertainty of the average rock density of 2\,\%, which is equivalent to an energy loss
uncertainty of 0.4~\GeV{} in the vertical direction.

Good agreement between the muon energy-loss 
calculation used here~\cite{soft:GEANT,soft:GEANTpatch} and other 
approaches~\cite{eloss:lo,eloss:groom,eloss:mmc2} 
is found. The residual differences correspond to a momentum scale uncertainty below or less than 0.3\,\% in 
the vertical direction.

\begin{figure*}[t!]
\centering
\centering
\includegraphics[width=\linewidth]{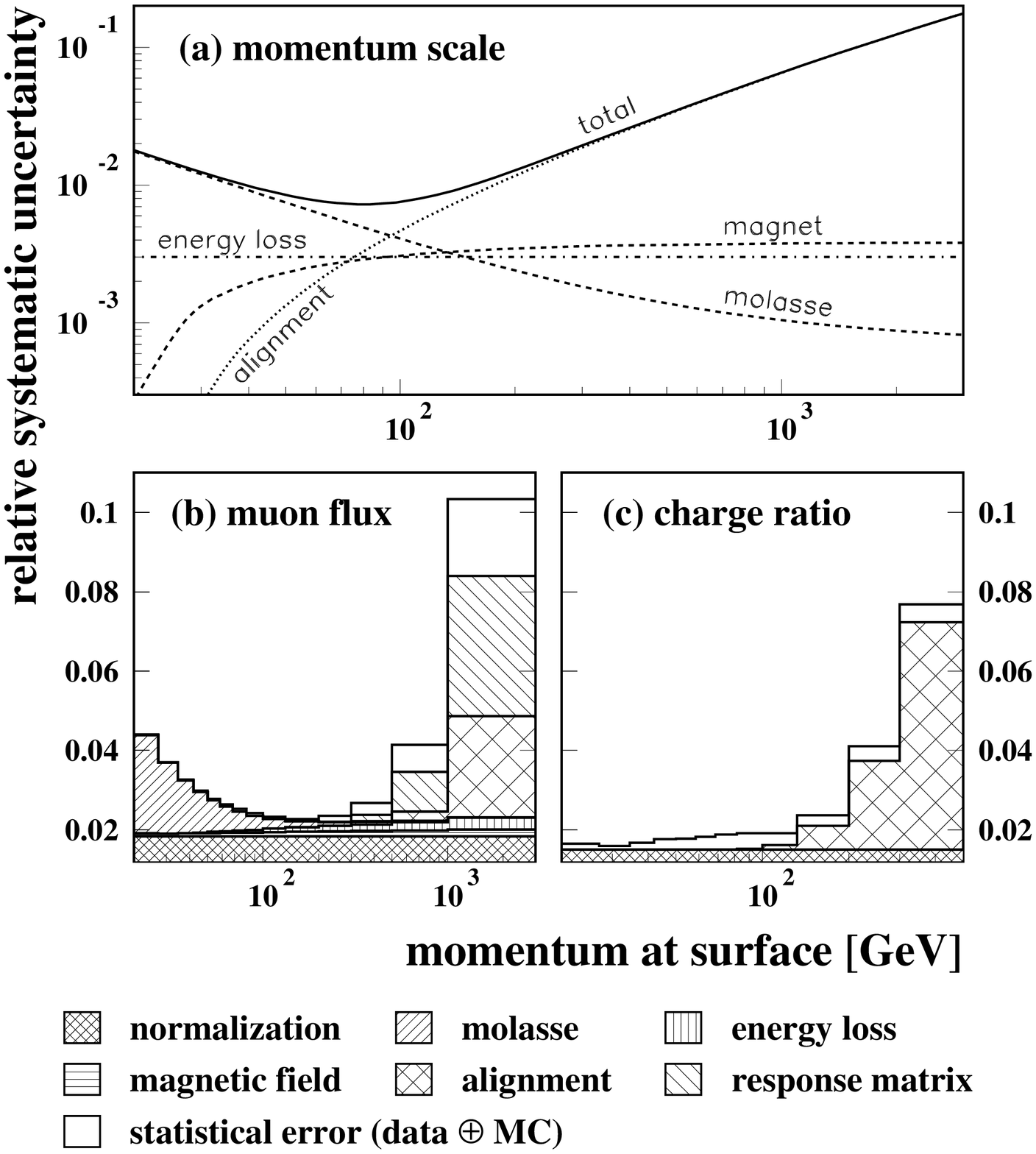}
\label{fig:crats}
\caption{Relative uncertainties of the vertical zenith angle bin measurements for (a) the momentum scale, (b) the
       muon flux and (c) the charge ratio. 
       The individual contributions are added in quadrature.}
\label{fig:scsys}
\end{figure*}

The relative momentum scale uncertainties for vertically incident muons are displayed in 
Figure~\ref{fig:scsys}(a). At low energies the molasse uncertainty
contributes the most, whereas above 100~\GeV{} the alignment uncertainties dominate. 

\subsection*{Detector matrix uncertainty}
The limited Monte Carlo statistics affects the precision of the detector matrix $\mathbf D$. 
Below 200~\GeV{}, it dominates the 
total statistical uncertainty in the denominator of Equation~(\ref{eq:lsqchi2}), 
contributing about 0.5\,\% to the total uncertainty per zenith angle bin.

In order to estimate the influence of the uncertainty of the momentum resolution on the measured 
muon flux, the minimization of Equation~(\ref{eq:lsqchi2}) is repeated with different detector
matrices, for which the momentum
resolution is altered by $\pm$8\,\%. This corresponds to the estimated uncertainty of its Monte Carlo prediction. 
As expected, no differences are found at low momenta. Above 200~\GeV{}, 
the observed relative flux change $\Delta_\Phi$ is well described by
\begin{equation}
    \Delta_\Phi=
    c\,\cdot (p-0.2 \,\mathrm{\TeV{}}),
\end{equation}
with $c=0.03 \,\mathrm{\TeV{}}^{-1}$. The observed difference between the high-energy muon flux
measured in different detector regions leads to a somewhat larger value of 
$c=0.06 \,\mathrm{\TeV{}}^{-1}$.
    %
    %
\subsection*{Total uncertainty}
The total uncertainties of the muon flux and charge ratio are obtained by adding the individual
 contributions in quadrature.
The different sources of the vertical uncertainties are shown in Figures~\ref{fig:scsys}(b) 
and (c).
The muon flux uncertainty is dominated by the uncertainty of the molasse overburden at low momenta
and by the alignment and resolution uncertainty at high momenta. The minimal uncertainty is 2.3\,\% at 150~\GeV{} 
in the vertical direction. 
The vertical
charge ratio uncertainty is below 2\,\% up to momenta of 100~\GeV{}. Above this momentum, it  
rises rapidly with the alignment uncertainties.

These uncertainties are fully correlated between different momenta for a given zenith angle bin.
As approximately the same detector parts are used to measure the muons in neighboring zenith angles,  
the systematic uncertainties are also correlated with respect to the zenith angle. The estimated
correlation coefficients are listed in Table~\ref{tab:zsyscorr}. 
    %
    %
\subsection*{Z-events}
\begin{figure}[t!]
   \begin{center}
       \includegraphics[width=\linewidth] {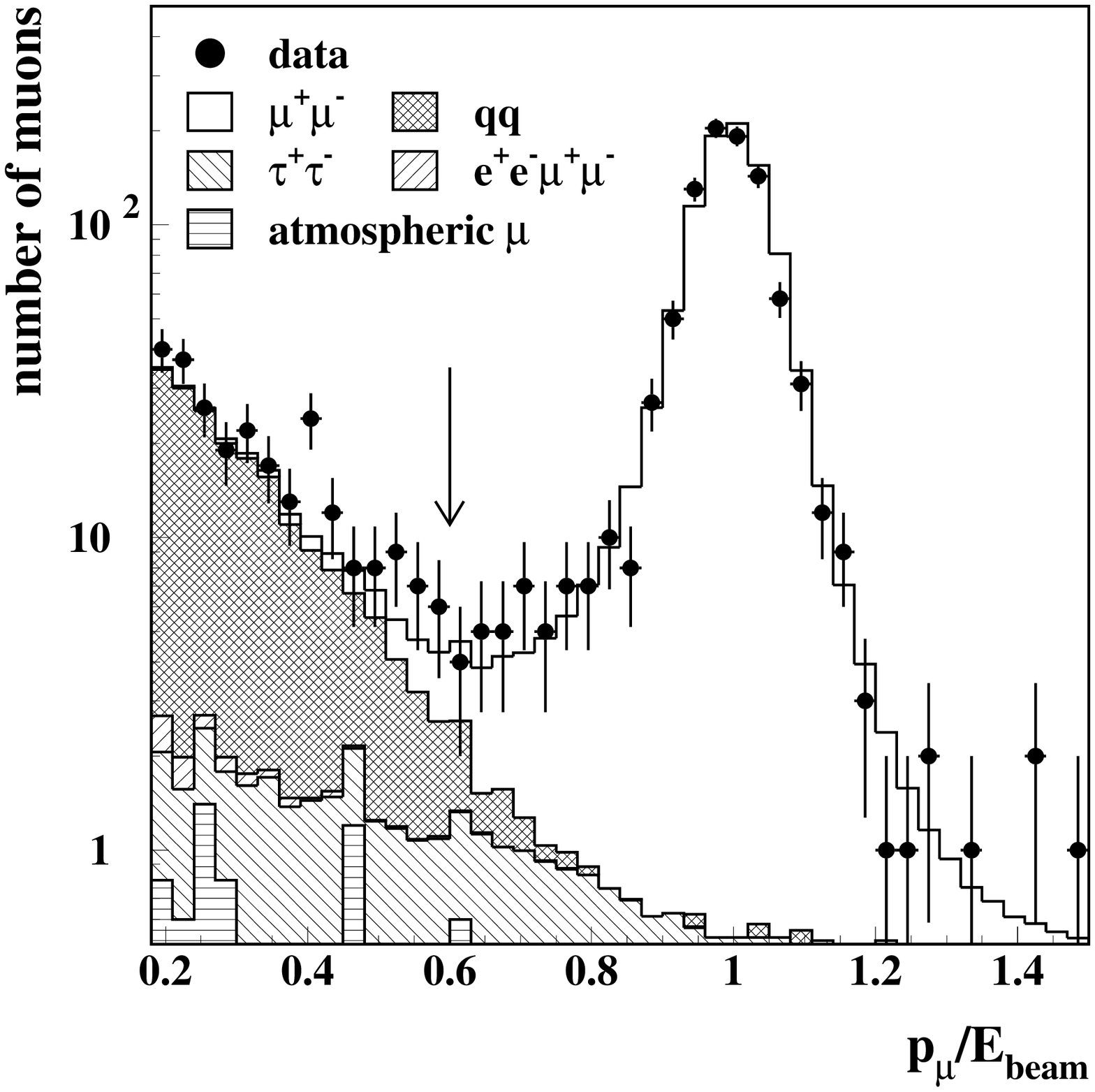}
        \caption[Z momentum distribution]{Momentum distribution of the 
	selected Z-events and the background. The Monte Carlo~\cite{zs:koralz,zs:pythia,zs:lep4f} 
	events are normalized to the Standard Model expectation as given in 
Equation~(\ref{eq:smexp}). 
        The arrow indicates the low momentum cut.}
        \label{fig:z0e}
   \end{center}
\end{figure}

The understanding of the detector is validated by analyzing the muons produced at LEP via the process
\begin{equation*}
    \epem \rightarrow \Zo \rightarrow \mm\; ,
\end{equation*}
recorded during the  
LEP calibration runs at a mean centre-of-mass energy of 91.27~\GeV{}. The selection criteria include 
the requirement of a muon track close to the collision point and an event-time in coincidence with the
LEP beam crossing time. The number of selected muons with a momentum above 60\,\% of the beam
energy is converted to an absolute cross section resulting in
\begin{eqnarray}
\sigma_{\mu^+\mu^-}^\text{L3+C} \, = \,
1.447 \pm 0.071\;(\text{stat.}) \,\pm \, 0.021\;(\text{syst.}) \;\text{nb} \; .
\end{eqnarray}
Here the quoted systematic uncertainty includes only
sources which are not relevant to the muon spectrum measurement, such as the 
luminosity.\\
Using the LEP precision measurements~\cite{zs:pdg}, the Standard Model
expectation of $\sigma_{\mu^+\mu^-}$ is calculated~\cite{zs:zfitnew} to be
\begin{equation}
\sigma_{\mu^+\mu^-}^\text{SM}=1.4840 \pm 0.0013 \;\text{nb}\,,
\label{eq:smexp}
\end{equation}
which is in excellent agreement to the value measured here.\\
Thus this study verifies the L3+C acceptance calculation and a
normalization uncertainty of $<$5.2\% at 68\% C.L. can be stated. Although this number is larger than the 
estimated systematic uncertainty of the muon flux normalization, it provides
an absolute systematic cross-check qualitatively different from the relative
studies described above.\\
The momentum distribution of 
the selected events, displayed in Figure~\ref{fig:z0e}, shows good agreement between the data and the simulation. From the peak position of the data, 
an absolute momentum scale uncertainty of $<$\,370~\MeV{} and a single octant
alignment uncertainty of $<$\,0.1~\TeV{}$^{-1}$ is derived.\\
As can be seen in figure Figure~\ref{fig:reso}(a), the momentum resolution derived from the $Z$ muon
sample agrees well with the one measured with atmospheric muons.

%
%
\section*{Results}
The muon fluxes, $\Phi$, conventionally multiplied by the third power of the momentum, and the charge ratios, $R$, 
are listed for each zenith angle bin in Tables 
\ref{tab:flux10}$-$\ref{tab:flux03} with their statistical and systematic uncertainties.
The statistical correlation coefficients, $\rho$, between neighboring momentum bins, as derived from 
Equation~(\ref{eq:Vm}), are also given. Due to the
limited detector resolution these correlations are inevitable. However, the momentum
binning  is chosen such that only neighboring bins have a significant correlation.  

The average momenta, $\langle p \rangle$, within a momentum range $[p_1,p_2]$
 are calculated ~\cite{meth:pmean} by fitting the phenomenological
muon flux function from Reference~\cite{cosmu:thct} to our data and solving
\begin{equation}
    \Phi(\langle p \rangle ) = \frac{1}{p_2-p_1} \int_{p_1}^{p_2} \Phi(p)\,dp \, \, .
\end{equation}

It should be noted that the fluxes are neither corrected for the altitude of L3+C nor for 
the atmospheric profile 
\begin{figure}[t!]
   \begin{center}
       \includegraphics[width=0.93\linewidth]{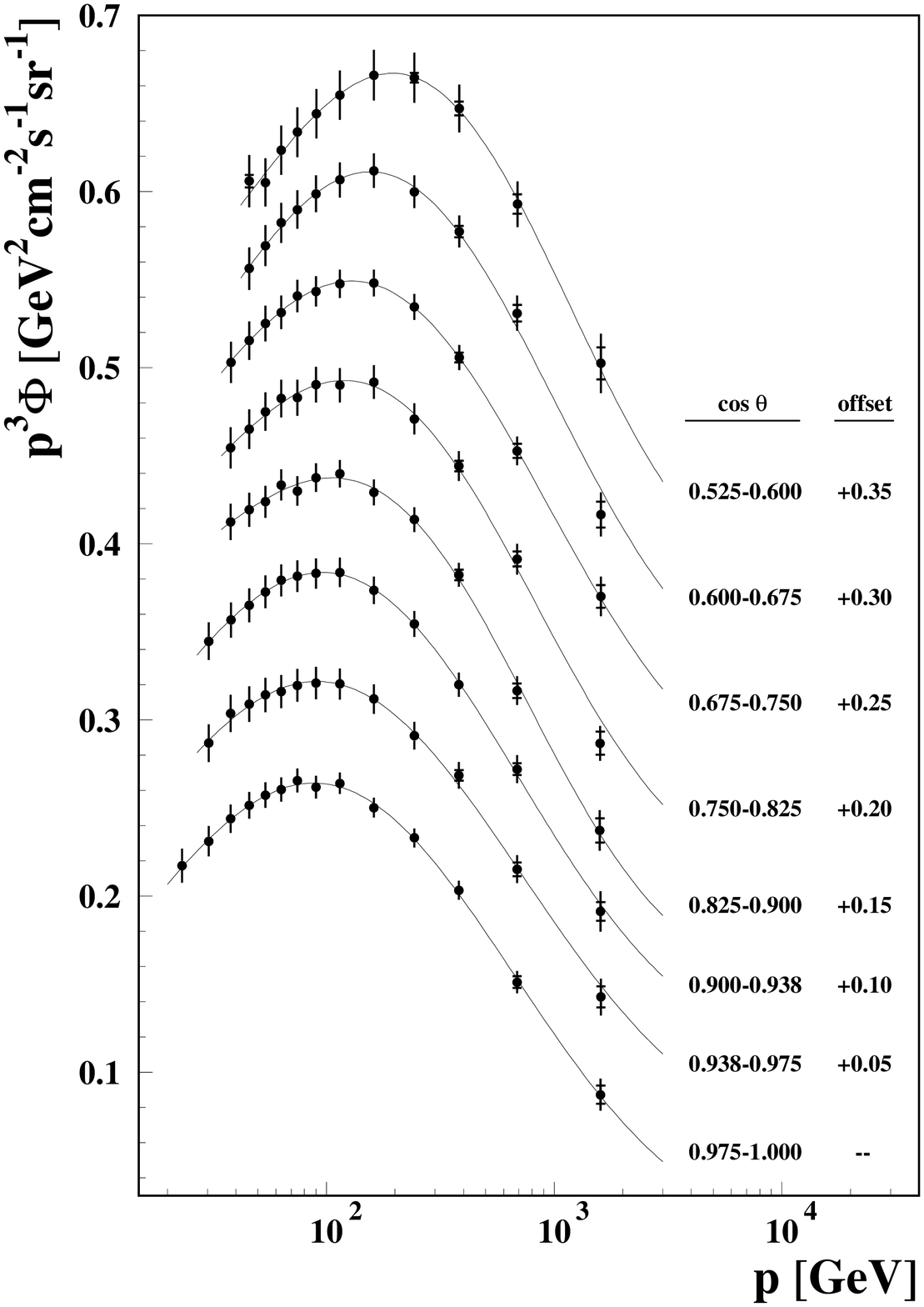}
        \caption{The measured muon flux for zenith angles ranging from 0$^\circ$~(bottom) 
                 to 58$^\circ$~(top). 
                  The inner bars denote the statistical
                  uncertainty, the full bars show the total uncertainty. For better visibility,
                  an offset of 0.05~\GeV{}$^2$cm$^{-2}$s$^{-1}$sr$^{-1}$ was added 
                  consecutively and lines are shown to guide the eye. }
        \label{fig:spectrum}
   \end{center}
\end{figure}
to avoid additional theoretical uncertainties. 
Instead, we quote the average atmospheric mass overburden $X$ above L3+C,
which  was continuously measured 
with balloon flights from close to the experiment 
to altitudes of over 30~km~\cite{l3c:atprof}. 
The parameterization of Reference~\cite{misc:atprof} is used to describe the mass profile $X$
in g\,cm$^{-2}$
as a function of the altitude $h$ in km above sea level:
\begin{equation}
    X(h)=
    \begin{cases}
        A\,(h_b-h)^{(\alpha+1)}, \quad h\le 11\\
        B\,e^{-\frac{h}{h_0}}, \qquad\qquad\,\,\, h > 11
    \end{cases}
\end{equation}
A fit to the live-time weighted balloon data yields 
$A=8.078 \times 10^{-5}$, $B=1332$, $h_b=39.17 $, $h_0=6.370$ and $\alpha=3.461$.

\begin{figure}[t!]
   \begin{center}
     \includegraphics[width=\linewidth]{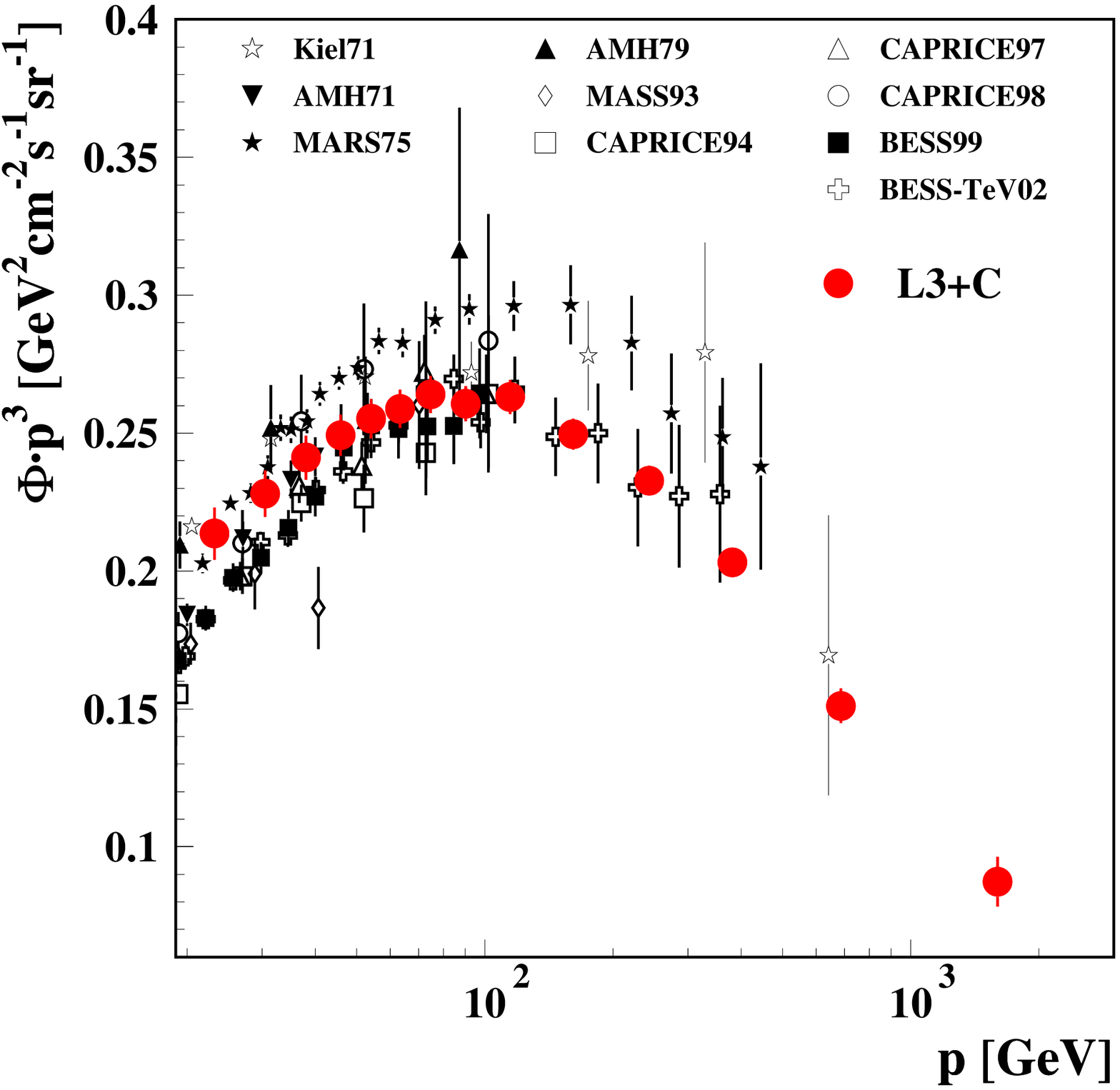}
     \caption[Vertical muon spectrum]{{The L3+C vertical muon spectrum compared to previous direct measurements 
                     providing an absolute flux normalization. 
                     All data are extra\-polated to sea level.}}
   \label{fig:vertflux}
   \end{center}
\end{figure}

The measured muon fluxes at the L3+C altitude are shown in 
Figure~\ref{fig:spectrum} for each zenith angle bin.
As no previous continuous zenith angle measurements exist in the large energy range examined here, only 
the vertical flux can be compared to other experiments, as shown in  
Figure~\ref{fig:vertflux}. Only
measurements providing an absolute 
normalization~\cite{cosmu:alkof,cosmu:ayre,cosmu:bate,cosmu:green,cosmu:depascale,cosmu:CAPRICE,cosmu:CAPRICE98,cosmu:Bess99,cosmu:Bess02} 
are taken into account. 
The data are extrapolated to sea level using the muon flux predictions of the \texttt{TARGET}~\cite{soft:target} program.

The comparison to low energy experiments~\cite{cosmu:bate,cosmu:green,cosmu:depascale,cosmu:CAPRICE,cosmu:CAPRICE98,cosmu:Bess99} 
gives a good overall agreement with this analysis above about 40~\GeV{}.
At lower momenta, a systematic slope difference seems to be present, which corresponds to about three standard deviations
of the systematic molasse uncertainty estimated above.

Only three previous experiments measured a normalized spectrum at high energies.
The shape of the Kiel measurements~\cite{cosmu:alkof} 
agree with this result over the full momentum range, but a lower flux normalization is determined by L3+C.\\
The data obtained with the MARS apparatus~\cite{cosmu:ayre} significantly disagree with this result,
both in shape and normalization.\\
Above momenta of 50~\GeV, the recent muon flux measurement from 
BESS-TeV~\cite{cosmu:Bess02} is in very good agreement with this result. 

\begin{figure}[t!]
   \begin{center}
       \includegraphics[width=0.93\linewidth]{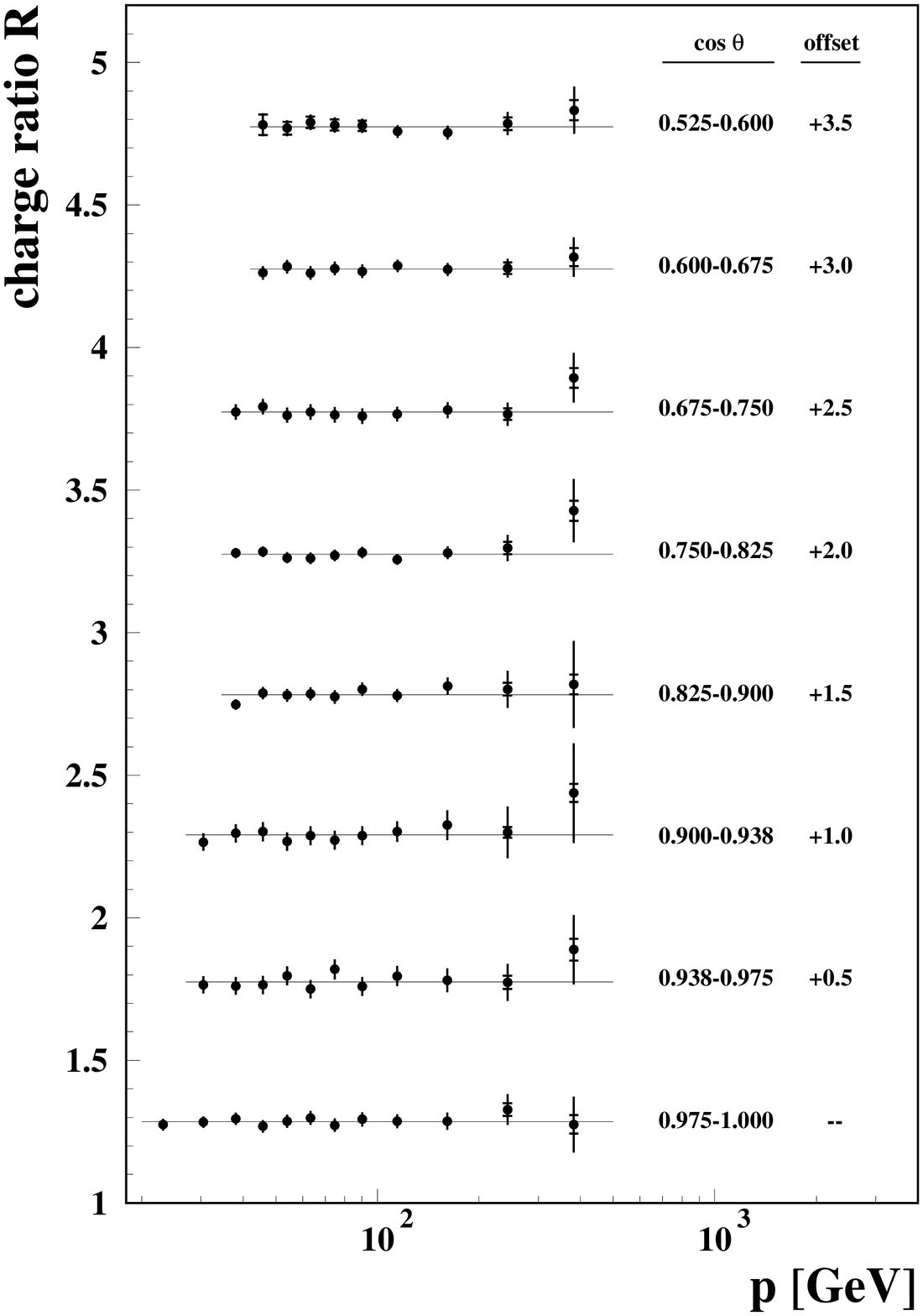}
        \caption{The measured muon charge ratio for zenith angles ranging from 0$^\circ$~(bottom) 
                 to 58$^\circ$~(top). 
                  The inner bars denote the statistical uncertainty,
                  the full bars show the total uncertainty. For better visibility,
                  an offset of 0.5 was added consecutively and lines are shown to guide the eye. }
        \label{fig:crat}
   \end{center}
\end{figure}

The measured charge ratios at the L3+C altitude are shown in 
Figure~\ref{fig:crat}  for each zenith angle bin up to momenta of 500~\GeV{}.
In the considered momentum range, the charge ratio is independent of the momentum within the experimental
uncertainties.
The mean value in the vertical direction is found to be
1.285 $\pm$ 0.003 (stat.) $\pm$ 0.019 (sys.) with a $\chi^2/\mathrm{ndf}=$9.5/11. This is in
good agreement with the average of all previous measurements, 1.270 $\pm$ 0.003 (stat.) $\pm$ 0.015 (sys.) 
~\cite{cosmu:thct}. It is worth noting, that the precision of the 
data of a single L3+C zenith angle bin is comparable to the combined uncertainty of all data collected
in the past.

%
%
\section*{Acknowledgments}
We wish to acknowledge the contribution of all the engineers and
technicians who have participated in the construction and the maintenance of the L3 and
L3+C experiments. Those of us who are not from member states thank CERN for its hospitality 
and help.
Furthermore, we would like to thank the Bartol group for providing the \texttt{TARGET} program.\\
%
%

\appendix
%
%
%
\newpage

\typeout{   }     
\typeout{Using author list for paper 287 -  }
\typeout{$Modified: Jul 15 2001 by smele $}
\typeout{!!!!  This should only be used with document option a4p!!!!}
\typeout{   }
%
%
%
%
%
%

\newcount\tutecount  \tutecount=0
\def\tutenum#1{\global\advance\tutecount by 1 \xdef#1{\the\tutecount}}
\def\tute#1{$^{#1}$}
\tutenum\aachen            
\tutenum\nikhef            
\tutenum\mich              
\tutenum\lapp              
\tutenum\basel             
\tutenum\lsu               
\tutenum\beijing           
\tutenum\itp               
\tutenum\hum               
\tutenum\bologna           
\tutenum\tata              
\tutenum\ne                
\tutenum\bucharest         
\tutenum\budapest          
\tutenum\mit               
\tutenum\panjab            
\tutenum\debrecen          
\tutenum\dublin            
\tutenum\florence          
\tutenum\cern              
\tutenum\wl                
\tutenum\geneva            
\tutenum\hamburg           
\tutenum\hefei             
\tutenum\lausanne          
\tutenum\lyon              
\tutenum\madrid            
\tutenum\florida           
\tutenum\milan             
\tutenum\moscow            
\tutenum\naples            
\tutenum\cyprus            
\tutenum\nymegen           
\tutenum\osaka             
\tutenum\caltech           
\tutenum\perugia           
\tutenum\peters            
\tutenum\cmu               
\tutenum\potenza           
\tutenum\prince            
\tutenum\riverside         
\tutenum\rome              
\tutenum\salerno           
\tutenum\ucsd              
\tutenum\santiago          
\tutenum\sofia             
\tutenum\korea             
\tutenum\taiwan            
\tutenum\tsinghua          
\tutenum\purdue            
\tutenum\psinst            
\tutenum\zeuthen           
\tutenum\eth               

{
\parskip=0pt
\noindent
{\bf The L3 Collaboration:}
\ifx\selectfont\undefined
 \baselineskip=10.8pt
 \baselineskip\baselinestretch\baselineskip
 \normalbaselineskip\baselineskip
 \ixpt
\else
 \fontsize{9}{10.8pt}\selectfont
\fi
\medskip
\tolerance=10000
\hbadness=5000
\raggedright
\hsize=162truemm\hoffset=0mm
\def\r{\rlap,}
\noindent

P.Achard\r\tute\geneva\ 
O.Adriani\r\tute{\florence}\ 
M.Aguilar-Benitez\r\tute\madrid\
M.van~den~Akker\r\tute\nymegen\ 
J.Alcaraz\r\tute{\madrid}\ 
G.Alemanni\r\tute\lausanne\
J.Allaby\r\tute\cern\
A.Aloisio\r\tute\naples\ 
M.G.Alviggi\r\tute\naples\
H.Anderhub\r\tute\eth\
% ...

\newpage
\begin{list}{A}{\itemsep=0pt plus 0pt minus 0pt\parsep=0pt plus 0pt minus 0pt
                \topsep=0pt plus 0pt minus 0pt}
\item[\aachen]
 III. Physikalisches Institut, RWTH, D-52056 Aachen, Germany$^{\S}$
\item[\nikhef] National Institute for High Energy Physics, NIKHEF, 
     and University of Amsterdam, NL-1009 DB Amsterdam, The Netherlands
\item[\mich] University of Michigan, Ann Arbor, MI 48109, USA
\item[\lapp] Laboratoire d'Annecy-le-Vieux de Physique des Particules, 
     LAPP,IN2P3-CNRS, BP 110, F-74941 Annecy-le-Vieux CEDEX, France
\item[\basel] Institute of Physics, University of Basel, CH-4056 Basel,
     Switzerland
\item[\lsu] Louisiana State University, Baton Rouge, LA 70803, USA
\item[\beijing] Institute of High Energy Physics, IHEP, 
  100039 Beijing, China$^{\triangle}$ 
\item[$\itp$] ITP, Academia Sinica, 100039 Beijing, China 
\item[$\hum$] Humboldt University, D-10115 Berlin, Germany. 
\item[\bologna] University of Bologna and INFN-Sezione di Bologna, 
     I-40126 Bologna, Italy
\item[\tata] Tata Institute of Fundamental Research, Mumbai (Bombay) 400 005, India
\item[\ne] Northeastern University, Boston, MA 02115, USA
\item[\bucharest] Institute of Atomic Physics and University of Bucharest,
     R-76900 Bucharest, Romania
\item[\budapest] Central Research Institute for Physics of the 
     Hungarian Academy of Sciences, H-1525 Budapest 114, Hungary$^{\ddag}$
\item[\mit] Massachusetts Institute of Technology, Cambridge, MA 02139, USA
\item[\panjab] Panjab University, Chandigarh 160 014, India
\item[\debrecen] KLTE-ATOMKI, H-4010 Debrecen, Hungary$^\P$
\item[\dublin] Department of Experimental Physics,
  University College Dublin, Belfield, Dublin 4, Ireland
\item[\florence] University of Florence and INFN, Sezione di Firenze,  
     I-50019 Sesto Fiorentino, Italy
\item[\cern] European Laboratory for Particle Physics, CERN, 
     CH-1211 Geneva 23, Switzerland
\item[\wl] World Laboratory, FBLJA  Project, CH-1211 Geneva 23, Switzerland
\item[\geneva] University of Geneva, CH-1211 Geneva 4, Switzerland
\item[\hamburg] University of Hamburg, D-22761 Hamburg, Germany
\item[\hefei] Chinese University of Science and Technology, USTC,
      Hefei, Anhui 230 029, China$^{\triangle}$
\item[\lausanne] University of Lausanne, CH-1015 Lausanne, Switzerland
\item[\lyon] Institut de Physique Nucl\'eaire de Lyon, 
     IN2P3-CNRS,Universit\'e Claude Bernard, 
     F-69622 Villeurbanne, France
\item[\madrid] Centro de Investigaciones Energ{\'e}ticas, 
     Medioambientales y Tecnol\'ogicas, CIEMAT, E-28040 Madrid,
     Spain${\flat}$ 
\item[\florida] Florida Institute of Technology, Melbourne, FL 32901, USA
\item[\milan] INFN-Sezione di Milano, I-20133 Milan, Italy
\item[\moscow] Institute of Theoretical and Experimental Physics, ITEP, 
     Moscow, Russia
\item[\naples] INFN-Sezione di Napoli and University of Naples, 
     I-80125 Naples, Italy
\item[\cyprus] Department of Physics, University of Cyprus,
     Nicosia, Cyprus
\item[\nymegen] University of Nijmegen and NIKHEF, 
     NL-6525 ED Nijmegen, The Netherlands
\item[$\osaka$] Osaka City University, Osaka 558-8585, Japan     
\item[\caltech] California Institute of Technology, Pasadena, CA 91125, USA
\item[\perugia] INFN-Sezione di Perugia and Universit\`a Degli 
     Studi di Perugia, I-06100 Perugia, Italy        
\item[\peters] Nuclear Physics Institute, St. Petersburg, Russia
\item[\cmu] Carnegie Mellon University, Pittsburgh, PA 15213, USA
\item[\potenza] INFN-Sezione di Napoli and University of Potenza, 
     I-85100 Potenza, Italy
\item[\prince] Princeton University, Princeton, NJ 08544, USA
\item[\riverside] University of California, Riverside, CA 92521, USA
\item[\rome] INFN-Sezione di Roma and University of Rome, ``La Sapienza",
     I-00185 Rome, Italy
\item[\salerno] University and INFN, Salerno, I-84100 Salerno, Italy
\item[\ucsd] University of California, San Diego, CA 92093, USA
\item[$\santiago$] University of Santiago de Compostela, E-15706 Santiago, Spain
\item[\sofia] Bulgarian Academy of Sciences, Central Lab.~of 
     Mechatronics and Instrumentation, BU-1113 Sofia, Bulgaria
\item[\korea]  The Center for High Energy Physics, 
     Kyungpook National University, 702-701 Taegu, Republic of Korea
\item[\taiwan] National Central University, Chung-Li, Taiwan, China
\item[\tsinghua] Department of Physics, National Tsing Hua University,
      Taiwan, China
\item[\purdue] Purdue University, West Lafayette, IN 47907, USA
\item[\psinst] Paul Scherrer Institut, PSI, CH-5232 Villigen, Switzerland
\item[\zeuthen] DESY, D-15738 Zeuthen, Germany
\item[\eth] Eidgen\"ossische Technische Hochschule, ETH Z\"urich,
     CH-8093 Z\"urich, Switzerland

\item[\S]  Supported by the German Bundesministerium 
        f\"ur Bildung, Wissenschaft, Forschung und Technologie.
\item[\ddag] Supported by the Hungarian OTKA fund under contract
numbers T019181, F023259 and T037350.
\item[\P] Also supported by the Hungarian OTKA fund under contract
  number T026178.
\item[$\flat$] Supported also by the Comisi\'on Interministerial de Ciencia y 
        Tecnolog{\'\i}a.
\item[$\sharp$] Also supported by CONICET and Universidad Nacional de La Plata,
        CC 67, 1900 La Plata, Argentina.
\item[$\triangle$] Supported by the National Natural Science
  Foundation of China.
\item[$\odot$] On leave from the Moscow Physical Engineering Institute (MePhl).
\item[$\diamond$] On leave from JINR, RU-141980 Dubna, Russia.
\item[$\dagger$] Deceased

\end{list}
}
\vfill


\newpage

%
%
\section*{}

\begin{table*}[!ht]
  \begin{center}
\small
     \begin{tabular}{|c||c ccccccc|}\hline
 &0.525-&0.600-&0.675-&0.750-&0.825-&0.900-&0.938-&0.975- \\
\thi{$\cos \theta $} &0.600&0.675&0.750&0.825&0.900&0.938&0.975&1.000 \\\hline\hline
0.525-0.600 &  1.00 & 0.91 & 0.99 & 0.94 & 0.93 & 0.00 & 0.00 & 0.00\\
0.600-0.675 &  0.91 & 1.00 & 0.96 & 0.76 & 0.88 & 0.00 & 0.00 & 0.00\\
0.675-0.750 &  0.99 & 0.96 & 1.00 & 0.91 & 0.95 & 0.00 & 0.00 & 0.00\\
0.750-0.825 &  0.94 & 0.76 & 0.91 & 1.00 & 0.93 & 0.00 & 0.00 & 0.00\\
0.825-0.900 &  0.93 & 0.88 & 0.95 & 0.93 & 1.00 & 0.01 & 0.01 & 0.01\\
0.900-0.938 &  0.00 & 0.00 & 0.00 & 0.00 & 0.01 & 1.00 & 1.00 & 1.00\\
0.938-0.975 &  0.00 & 0.00 & 0.00 & 0.00 & 0.01 & 1.00 & 1.00 & 1.00\\
0.975-1.000 &  0.00 & 0.00 & 0.00 & 0.00 & 0.01 & 1.00 & 1.00 & 1.00\\ \hline
    \end{tabular}	
        \caption{Correlation coefficients of the detector-related systematic uncertainties between
different zenith angle bins from 0$^{\circ}$ to 58$^{\circ}$}.
        \label{tab:zsyscorr}
  \end{center}
\end{table*}	
     

\begin{table*}[!t]
  \begin{center}
     \scriptsize
     \begin{tabular}{|r@{-}l| c || c | c | c | c || c | c | c | c |}\hline
         \multicolumn{2}{|c|}{}\fluxtabHone               \\
         \fluxtabFone &\fluxtabRone \\
         \fluxtabFtwo &\fluxtabRtwo \\
         \fluxtabHtwo \fluxtabHone  \\\hline\hline

   20.0&27.0  &  23.18&  0.217&    0.4&\tlo{  -0.24}&    4.4&  1.274&    0.7&\tlo{  -0.24}&    1.5\\
   27.0&34.5  &  30.47&  0.231&    0.3&\tlo{  -0.22}&    3.7&  1.284&    0.6&\tlo{  -0.22}&    1.5\\
   34.5&42.0  &  38.02&  0.244&    0.4&\tlo{  -0.24}&    3.3&  1.295&    0.8&\tlo{  -0.24}&    1.5\\
   42.0&50.0  &  45.78&  0.252&    0.5&\tlo{  -0.26}&    3.0&  1.269&    0.9&\tlo{  -0.26}&    1.5\\
   50.0&58.5  &  54.04&  0.257&    0.5&\tlo{  -0.31}&    2.8&  1.286&    1.0&\tlo{  -0.31}&    1.5\\
   58.5&68.5  &  63.25&  0.261&    0.5&\tlo{  -0.31}&    2.6&  1.298&    1.0&\tlo{  -0.31}&    1.5\\
   68.5&81.5  &  74.63&  0.266&    0.5&\tlo{  -0.30}&    2.5&  1.273&    1.1&\tlo{  -0.29}&    1.5\\
   81.5&100   &  90.13&  0.262&    0.6&\tlo{  -0.26}&    2.4&  1.293&    1.2&\tlo{  -0.26}&    1.5\\
    100&132   & 114.5 &  0.264&    0.5&\tlo{  -0.22}&    2.3&  1.286&    1.0&\tlo{  -0.21}&    1.6\\
    132&200   & 161.3 &  0.250&    0.5&\tlo{  -0.22}&    2.2&  1.287&    1.1&\tlo{  -0.22}&    2.1\\
    200&300   & 243.0 &  0.233&    0.8&\tlo{  -0.27}&    2.2&  1.327&    1.7&\tlo{  -0.27}&    3.7\\
    300&500   & 381.9 &  0.203&    1.2&\tlo{  -0.28}&    2.4&  1.276&    2.6&&    7.2\\ \cline{8-11}
    500&1000  & 687.2 &  0.151&    2.3&\tlo{  -0.30}&    3.5  \\
   1000&3000  &1599   &  0.087&    6.0&             &    8.4  \\\cline{1-7}

  \end{tabular}	
        \caption{Muon flux, $\Phi$, multiplied with the third power of the momentum, 
	and charge ratio for 0.975$\,<$ cos$\,\mathrm{\theta}<\,$1.000. The statistical, $\rm \Delta^{stat}$,
	and the systematical, $\rm \Delta^{syst}$, uncertainties are given.
	$\rho_{\Phi}$ and $\rho_{R}$ are the 
	statistical correlation coefficients between neighboring momentum bins, as derived from Equation~(\ref{eq:Vm}).}
        \label{tab:flux10}
  \end{center}
\end{table*}	

\begin{table*}[!ht]
  \begin{center}
     \scriptsize
     \begin{tabular}{|r@{-}l| c || c | c | c | c || c | c | c | c |}\hline
         \multicolumn{2}{|c|}{}\fluxtabHone               \\
         \fluxtabFone &\fluxtabRone \\
         \fluxtabFtwo &\fluxtabRtwo \\
         \fluxtabHtwo \fluxtabHone  \\\hline\hline

   27.0&34.5  &  30.47&  0.237&    0.3&\tlo{  -0.23}&    4.5&  1.265&    0.6&\tlo{  -0.22}&    2.3\\
   34.5&42.0  &  38.02&  0.254&    0.4&\tlo{  -0.24}&    4.1&  1.261&    0.8&\tlo{  -0.24}&    2.3\\
   42.0&50.0  &  45.78&  0.259&    0.5&\tlo{  -0.26}&    3.8&  1.265&    1.0&\tlo{  -0.26}&    2.3\\
   50.0&58.5  &  54.04&  0.264&    0.5&\tlo{  -0.30}&    3.7&  1.297&    1.1&\tlo{  -0.30}&    2.3\\
   58.5&68.5  &  63.25&  0.266&    0.6&\tlo{  -0.31}&    3.5&  1.250&    1.1&\tlo{  -0.31}&    2.3\\
   68.5&81.5  &  74.63&  0.270&    0.6&\tlo{  -0.29}&    3.4&  1.319&    1.2&\tlo{  -0.29}&    2.3\\
   81.5&100   &  90.13&  0.271&    0.6&\tlo{  -0.25}&    3.3&  1.259&    1.3&\tlo{  -0.25}&    2.3\\
    100&132   & 114.5 &  0.270&    0.5&\tlo{  -0.21}&    3.2&  1.296&    1.1&\tlo{  -0.21}&    2.5\\
    132&200   & 161.3 &  0.262&    0.6&\tlo{  -0.21}&    3.1&  1.281&    1.2&\tlo{  -0.21}&    3.0\\
    200&300   & 243.0 &  0.241&    0.9&\tlo{  -0.27}&    3.1&  1.273&    1.8&\tlo{  -0.27}&    4.8\\
    300&500   & 382.1 &  0.219&    1.3&\tlo{  -0.27}&    3.2&  1.389&    2.7&&    8.3\\\cline{8-11}
    500&1000  & 687.8 &  0.165&    2.4&\tlo{  -0.30}&    4.2 \\
   1000&3000  & 1604  &  0.093&    6.5&             &    9.2 \\\cline{1-7}

   \end{tabular}	
        \caption{Muon flux and charge ratio for 0.938$\,<$ cos$\,\mathrm{\theta}<\,$0.975.}
        \label{tab:flux09}
  \end{center}
\end{table*}	

\begin{table*}[!ht]
  \begin{center}
     \scriptsize
     \begin{tabular}{|r@{-}l| c || c | c | c | c || c | c | c | c |}\hline
         \multicolumn{2}{|c|}{}\fluxtabHone               \\
         \fluxtabFone &\fluxtabRone \\
         \fluxtabFtwo &\fluxtabRtwo \\
         \fluxtabHtwo \fluxtabHone  \\\hline\hline

   27.0&34.5 &  30.47&  0.245&    0.4&\tlo{  -0.24}&    4.3&  1.265&    0.7&\tlo{  -0.24}&    2.3\\
   34.5&42.0 &  38.02&  0.257&    0.4&\tlo{  -0.26}&    3.9&  1.296&    0.9&\tlo{  -0.26}&    2.3\\
   42.0&50.0 &  45.78&  0.265&    0.5&\tlo{  -0.28}&    3.6&  1.302&    1.1&\tlo{  -0.28}&    2.3\\
   50.0&58.5 &  54.04&  0.273&    0.5&\tlo{  -0.31}&    3.4&  1.267&    1.1&\tlo{  -0.31}&    2.3\\
   58.5&68.5 &  63.25&  0.279&    0.5&\tlo{  -0.31}&    3.2&  1.287&    1.1&\tlo{  -0.30}&    2.3\\
   68.5&81.5 &  74.64&  0.282&    0.5&\tlo{  -0.27}&    3.1&  1.272&    1.1&\tlo{  -0.27}&    2.3\\
   81.5&100  &  90.14&  0.283&    0.5&\tlo{  -0.23}&    3.0&  1.288&    1.1&\tlo{  -0.23}&    2.4\\
    100&132  & 114.5 &  0.284&    0.5&\tlo{  -0.19}&    2.9&  1.303&    0.9&\tlo{  -0.19}&    2.6\\
    132&200  & 161.3 &  0.274&    0.5&\tlo{  -0.19}&    2.8&  1.325&    1.0&\tlo{  -0.19}&    3.8\\
    200&300  & 243.0 &  0.255&    0.7&\tlo{  -0.23}&    2.8&  1.300&    1.4&\tlo{  -0.23}&    6.8\\
    300&500  & 382.0 &  0.220&    1.1&\tlo{  -0.25}&    2.9&  1.437&    2.2&&   12.0\\\cline{8-11}
    500&1000 & 687.3 &  0.172&    1.9&\tlo{  -0.29}&    4.2  \\ 
   1000&3000 & 1599  &  0.091&    5.7&             &   11.3  \\\cline{1-7}

     \end{tabular}	
        \caption{Muon flux and charge ratio for 0.900$\,<$ cos$\,\mathrm{\theta}<\,$0.938.}
        \label{tab:flux08}
  \end{center}
\end{table*}	

\begin{table*}[!ht]
  \begin{center}
     \scriptsize
     \begin{tabular}{|r@{-}l| c || c | c | c | c || c | c | c | c |}\hline
         \multicolumn{2}{|c|}{}\fluxtabHone               \\
         \fluxtabFone &\fluxtabRone \\
         \fluxtabFtwo &\fluxtabRtwo \\
         \fluxtabHtwo \fluxtabHone  \\\hline\hline

   34.5&42.0 &  38.02&  0.263&    0.4&\tlo{  -0.28}&    3.9&  1.248&    0.8&\tlo{  -0.28}&    1.3\\
   42.0&50.0 &  45.78&  0.269&    0.5&\tlo{  -0.29}&    3.5&  1.288&    1.1&\tlo{  -0.29}&    1.3\\
   50.0&58.5 &  54.04&  0.274&    0.5&\tlo{  -0.32}&    3.3&  1.280&    1.1&\tlo{  -0.32}&    1.3\\
   58.5&68.5 &  63.25&  0.284&    0.6&\tlo{  -0.31}&    3.1&  1.285&    1.2&\tlo{  -0.31}&    1.3\\
   68.5&81.5 &  74.64&  0.280&    0.6&\tlo{  -0.28}&    2.9&  1.274&    1.2&\tlo{  -0.28}&    1.3\\
   81.5&100  &  90.14&  0.288&    0.6&\tlo{  -0.24}&    2.8&  1.302&    1.3&\tlo{  -0.24}&    1.3\\
    100&132  & 114.5 &  0.290&    0.5&\tlo{  -0.20}&    2.6&  1.280&    1.1&\tlo{  -0.20}&    1.4\\
    132&200  & 161.3 &  0.279&    0.6&\tlo{  -0.20}&    2.6&  1.313&    1.1&\tlo{  -0.20}&    2.0\\
    200&300  & 243.1 &  0.264&    0.8&\tlo{  -0.27}&    2.5&  1.301&    1.7&\tlo{  -0.27}&    4.7\\
    300&500  & 382.1 &  0.232&    1.3&\tlo{  -0.30}&    2.6&  1.318&    2.6&&   11.3\\\cline{8-11}
    500&1000 & 686.8 &  0.167&    2.5&\tlo{  -0.37}&    4.2 \\
   1000&3000 & 1587  &  0.087&    7.9&             &   10.4\\\cline{1-7}

    \end{tabular}	
        \caption{Muon flux and charge ratio for 0.825$\,<$ cos$\,\mathrm{\theta}<\,$0.900.}
        \label{tab:flux07}
  \end{center}
\end{table*}	

\begin{table*}[!ht]
  \begin{center}
     \scriptsize
     \begin{tabular}{|r@{-}l| c || c | c | c | c || c | c | c | c |}\hline
        \multicolumn{2}{|c|}{}\fluxtabHone               \\
        \fluxtabFone &\fluxtabRone \\
        \fluxtabFtwo &\fluxtabRtwo \\
        \fluxtabHtwo \fluxtabHone  \\\hline\hline

   34.5&42.0 &  38.02 &  0.255&    0.5&\tlo{  -0.32}&    4.6&  1.279&    1.0&\tlo{  -0.32}&    1.0\\
   42.0&50.0 &  45.78 &  0.265&    0.5&\tlo{  -0.32}&    4.2&  1.284&    1.1&\tlo{  -0.32}&    1.0\\
   50.0&58.5 &  54.04 &  0.275&    0.6&\tlo{  -0.35}&    3.9&  1.263&    1.1&\tlo{  -0.35}&    1.0\\
   58.5&68.5 &  63.25 &  0.283&    0.6&\tlo{  -0.33}&    3.7&  1.261&    1.2&\tlo{  -0.33}&    1.0\\
   68.5&81.5 &  74.64 &  0.283&    0.6&\tlo{  -0.29}&    3.6&  1.271&    1.2&\tlo{  -0.29}&    1.0\\
   81.5&100  &  90.14 &  0.290&    0.6&\tlo{  -0.26}&    3.4&  1.281&    1.2&\tlo{  -0.26}&    1.0\\
    100&132  & 114.6  &  0.290&    0.5&\tlo{  -0.20}&    3.3&  1.256&    1.0&\tlo{  -0.20}&    1.1\\
    132&200  & 161.3  &  0.292&    0.5&\tlo{  -0.22}&    3.2&  1.280&    1.1&\tlo{  -0.22}&    1.4\\
    200&300  & 243.1  &  0.271&    0.8&\tlo{  -0.28}&    3.2&  1.297&    1.6&\tlo{  -0.28}&    3.1\\
    300&500  & 382.2  &  0.244&    1.2&\tlo{  -0.31}&    3.3&  1.428&    2.5&&    7.4\\\cline{8-11}
    500&1000 & 687.5  &  0.191&    2.2&\tlo{  -0.36}&    4.0\\
   1000&3000 &  1594  &  0.087&    7.6&             &    8.4\\\cline{1-7}

   \end{tabular}	
        \caption{ Muon flux and charge ratio for 0.750$\,<$ cos$\,\mathrm{\theta}<\,$0.825.}
        \label{tab:flux06}
  \end{center}
\end{table*}	

\begin{table*}[!ht]
  \begin{center}
     \scriptsize
     \begin{tabular}{|r@{-}l| c || c | c | c | c || c | c | c | c |}\hline
         \multicolumn{2}{|c|}{}\fluxtabHone               \\
         \fluxtabFone &\fluxtabRone \\
         \fluxtabFtwo &\fluxtabRtwo \\
         \fluxtabHtwo \fluxtabHone  \\\hline\hline

   34.5&42.0 &  38.02 &  0.253&    0.6&\tlo{  -0.38}&    4.6&  1.273&    1.2&\tlo{  -0.38}&    1.8\\
   42.0&50.0 &  45.78 &  0.265&    0.6&\tlo{  -0.35}&    4.1&  1.292&    1.1&\tlo{  -0.35}&    1.8\\
   50.0&58.5 &  54.04 &  0.275&    0.5&\tlo{  -0.38}&    3.7&  1.263&    1.1&\tlo{  -0.38}&    1.8\\
   58.5&68.5 &  63.25 &  0.281&    0.6&\tlo{  -0.34}&    3.4&  1.273&    1.2&\tlo{  -0.34}&    1.8\\
   68.5&81.5 &  74.64 &  0.291&    0.6&\tlo{  -0.31}&    3.1&  1.263&    1.2&\tlo{  -0.31}&    1.8\\
   81.5&100  &  90.15 &  0.293&    0.6&\tlo{  -0.27}&    2.9&  1.260&    1.2&\tlo{  -0.27}&    1.8\\
    100&132  & 114.6  &  0.298&    0.5&\tlo{  -0.22}&    2.7&  1.266&    1.0&\tlo{  -0.22}&    1.8\\
    132&200  & 161.4  &  0.298&    0.5&\tlo{  -0.23}&    2.5&  1.281&    1.0&\tlo{  -0.23}&    1.9\\
    200&300  & 243.2  &  0.285&    0.7&\tlo{  -0.29}&    2.4&  1.267&    1.6&\tlo{  -0.29}&    2.8\\
    300&500  & 382.3  &  0.256&    1.1&\tlo{  -0.31}&    2.6&  1.394&    2.5&&    5.7\\\cline{8-11}
    500&1000 & 688.2  &  0.203&    2.0&\tlo{  -0.35}&    3.5\\
   1000&3000 & 1601   &  0.120&    5.3&             &    7.8\\\cline{1-7}
\end{tabular}	
        \caption{ Muon flux and charge ratio for 0.675$\,<$ cos$\,\mathrm{\theta}<\,$0.750.}
        \label{tab:flux05}
  \end{center}
\end{table*}	

\begin{table*}[!ht]
  \begin{center}
     \scriptsize
     \begin{tabular}{|r@{-}l| c || c | c | c | c || c | c | c | c |}\hline
         \multicolumn{2}{|c|}{}\fluxtabHone               \\
         \fluxtabFone &\fluxtabRone \\
         \fluxtabFtwo &\fluxtabRtwo \\
         \fluxtabHtwo \fluxtabHone  \\\hline\hline
 
   42.0&50.0 &  45.79 &  0.256&    0.6&\tlo{  -0.41}&    4.7&  1.262&    1.3&\tlo{  -0.41}&    1.4\\
   50.0&58.5 &  54.05 &  0.269&    0.6&\tlo{  -0.42}&    4.3&  1.284&    1.2&\tlo{  -0.42}&    1.4\\
   58.5&68.5 &  63.26 &  0.282&    0.6&\tlo{  -0.36}&    4.0&  1.261&    1.3&\tlo{  -0.36}&    1.4\\
   68.5&81.5 &  74.65 &  0.290&    0.6&\tlo{  -0.31}&    3.7&  1.277&    1.3&\tlo{  -0.31}&    1.4\\
   81.5&100  &  90.16 &  0.299&    0.6&\tlo{  -0.26}&    3.5&  1.267&    1.2&\tlo{  -0.26}&    1.4\\
    100&132  & 114.6  &  0.307&    0.5&\tlo{  -0.22}&    3.3&  1.287&    1.0&\tlo{  -0.22}&    1.4\\
    132&200  & 161.4  &  0.312&    0.5&\tlo{  -0.24}&    3.1&  1.274&    1.0&\tlo{  -0.23}&    1.4\\
    200&300  & 243.2  &  0.300&    0.8&\tlo{  -0.32}&    3.0&  1.278&    1.6&\tlo{  -0.31}&    2.1\\
    300&500  & 382.5  &  0.277&    1.2&\tlo{  -0.35}&    3.1&  1.317&    2.4&&    4.7\\\cline{8-11}
    500&1000 & 688.5  &  0.231&    2.0&\tlo{  -0.38}&    3.8\\
   1000&3000 & 1602   &  0.117&    6.3&             &    8.7\\\cline{1-7}

\end{tabular}	
        \caption{ Muon flux and charge ratio for 0.600$\,<$ cos$\,\mathrm{\theta}<\,$0.675.}
        \label{tab:flux04}
  \end{center}
\end{table*}

	
\begin{table*}[!ht]
  \begin{center}
     \scriptsize
     \begin{tabular}{|r@{-}l| c || c | c | c | c || c | c | c | c |}\hline
         \multicolumn{2}{|c|}{}\fluxtabHone               \\
         \fluxtabFone &\fluxtabRone \\
         \fluxtabFtwo &\fluxtabRtwo \\
         \fluxtabHtwo \fluxtabHone  \\\hline\hline

   42.0&50.0 &  45.79&  0.256&    1.4&\tlo{  -0.49}&    5.6&  1.281&    2.8&\tlo{  -0.50}&    1.4\\
   50.0&58.5 &  54.05&  0.255&    0.9&\tlo{  -0.45}&    5.3&  1.269&    1.7&\tlo{  -0.45}&    1.4\\
   58.5&68.5 &  63.26&  0.273&    0.8&\tlo{  -0.38}&    5.1&  1.290&    1.6&\tlo{  -0.37}&    1.4\\
   68.5&81.5 &  74.65&  0.284&    0.8&\tlo{  -0.32}&    4.9&  1.280&    1.5&\tlo{  -0.32}&    1.4\\
   81.5&100  &  90.16&  0.294&    0.7&\tlo{  -0.25}&    4.7&  1.278&    1.4&\tlo{  -0.25}&    1.4\\
    100&132  & 114.6 &  0.305&    0.5&\tlo{  -0.22}&    4.6&  1.258&    1.1&\tlo{  -0.21}&    1.4\\
    132&200  & 161.5 &  0.316&    0.6&\tlo{  -0.22}&    4.5&  1.254&    1.2&\tlo{  -0.22}&    1.6\\
    200&300  & 243.4 &  0.315&    0.8&\tlo{  -0.30}&    4.4&  1.285&    1.7&\tlo{  -0.30}&    2.6\\
    300&500  & 382.7 &  0.297&    1.3&\tlo{  -0.33}&    4.4&  1.332&    2.6&&    5.6\\\cline{8-11}
    500&1000 & 689.3 &  0.243&    2.3&\tlo{  -0.36}&4.8\\
   1000&3000 &  1604 &  0.153&    5.9&             &9.4\\\cline{1-7}

    \end{tabular}	
        \caption{ Muon flux and charge ratio for 0.525$\,<$ cos$\,\mathrm{\theta}<\,$0.600.}
        \label{tab:flux03}
  \end{center}
\end{table*}	

\end{document}